\documentclass[preprint,showkeys,showpacs,prd]{revtex4}%
\usepackage{amsmath}
\usepackage{amsfonts}
\usepackage{amssymb}
\usepackage{graphicx}

\begin{document}
\def\be{\begin{equation}}
\def\ee{\end{equation}}
\def\bea{\begin{eqnarray}}
\def\eea{\end{eqnarray}}

\title{Multipolar Solutions}

%\classification{04.20.Jb; 95.30.Sf}


\author{Hernando Quevedo}
\affiliation{Instituto de Ciencias
Nucleares, Universidad Nacional Aut\'onoma de M\'exico\\
 AP 70543, M\'exico, DF 04510, Mexico \\ 
 ICRANet, Dipartimento di Fisica, Universit\'a di Roma ``La Sapienza'' \\  I-00185 Roma, Italy}

\begin{abstract}
A class of exact solutions of the Einstein-Maxwell equations is presented which contains
infinite sets of gravitoelectric, gravitomagnetic and electromagnetic multipole moments.
The multipolar structure of the solutions indicates that they can be used to describe the
exterior gravitational field of an arbitrarily rotating mass distribution endowed with 
an electromagnetic field. The presence of gravitational multipoles completely changes 
the structure of the spacetime because of the appearance of naked singularities in 
a confined spatial region. The possibility of covering this region with interior solutions
is analyzed in the case of a particular solution with quadrupole moment.
\end{abstract}

\maketitle

%%%%%%%%%%%%%%%%%%%%%%%%%%%%%%%%%%%%%%%%%%%%
%% MAINMATTER
%%%%%%%%%%%%%%%%%%%%%%%%%%%%%%%%%%%%%%%%%%%%

\section{Introduction}
\label{sec:int}

The problem of describing the gravitational field of astrophysical compact objects is very important in general relativity. 
Indeed, one of the most interesting applications of Einstein's general relativity consists in describing the exterior and
interior gravitational field generated by a specific mass distribution. Accordingly, it must be possible to find exact 
solutions of Einstein's field equations which correctly describe the influence of gravity inside and outside the mass 
distribution. 

This problem is solved in Newtonian gravity by finding the explicit form of the gravitational potential 
as a solution of the Poisson and the Laplace equations, respectively. To this end, it is usually assumed that the 
mass distribution satisfies certain symmetry conditions that are derived from observations. In particular, one can  
assume that the mass distribution does not depend on time and on the azimuthal angle, i.e. it is stationary and axisymmetric.
In this case, the expression of the gravitational potential can be written as an infinite series each term of which represents
a particular multipole moment. 

An additional important aspect is that all known astrophysical compact objects rotate with respect to 
observers situated at infinity where the gravitational field is negligible. In Newtonian gravity, however, it is well-known 
that there is no difference between the gravitational potentials of a static and of a rotating (stationary) mass distribution. 
The question arises whether in Einstein's theory of gravity one can find exact solutions that take into account the 
rotation and axial symmetry of the mass distribution. 
It
is therefore of importance and interest to describe the relativistic
gravitational fields of astrophysical compact objects in terms of
their multipole moments, in close analogy with the Newtonian theory,
taking into account their rotation and their internal structure.

In general relativity, the gravitational field of compact objects must be described by a metric $g_{\alpha\beta}\ (\alpha,\beta=0,1,2,3)$ 
satisfying Einstein's equations,
\be
R_{\alpha\beta} - \frac{1}{2} g_{\alpha\beta}R = 8\pi T_{\alpha\beta} \ ,
\ee
in the interior part of the object ($T_{\alpha\beta}\neq 0$) as well as outside in empty space $(T_{\alpha\beta}= 0$), where $T_{\alpha\beta}$
represents the energy-momentum tensor of the source of gravity.
Since at the classical level
 the most important sources of gravity are the mass distribution and the electromagnetic distribution, 
 it is clear that the problem of finding a metric that describes the gravitational 
 field of compact objects 
 can be divided into four related problems. The first one consists in finding an exact vacuum solution $(R_{\alpha\beta}=0$) that describes 
the field outside the mass distribution. Secondly, the electromagnetic distribution must be considered by solving the electrovacuum 
field equations
\be
R_{\alpha\beta}= 8\pi \left( F_{\alpha\gamma}F^{\ \gamma}_{\beta } - \frac{1}{4} g_{\alpha\beta} F_{\gamma\delta} F^{\gamma\delta}\right)\ ,
\ee
where $F_{\alpha\beta}$ is the Faraday tensor corresponding to a charge distribution $Q(x^\alpha)$.  
As for the interior region, it
is necessary to propose a model to describe the internal gravitational structure of the object.  
The most popular model is that of a perfect fluid 
\be
T_ {\alpha\beta} = (\rho + p )u_\alpha u_\beta - g_{\alpha\beta} p \ ,
\label{pf}
\ee
with density $\rho(x^\alpha)$, pressure $p(x^\alpha)$, and 4-velocity $u^\alpha$. Finally, the fourth part of the problem consists in considering the 
mass and charge distribution simultaneously, i.e. one needs an exact solution to the equations
\be
R_{\alpha\beta} - \frac{1}{2} g_{\alpha\beta}R = 8\pi \,  T_{\alpha\beta}(\rho, p,Q)\ ,
\ee
where a specific model must be proposed for the energy-momentum tensor. Each of the above problems is very difficult to solve because 
Einstein's field equations are highly non-linear. 

Soon after the
formulation of Einstein's theory of gravity, the first exterior solution with only a monopole
moment was discovered by Schwarzschild \cite{A9}. In 1917, Weyl
\cite{weyl} showed that the problem of finding static axisymmetric
vacuum solutions can generically be reduced to a single linear
differential equation whose general solution can be represented as
an infinite series. The explicit form of this solution resembles the
corresponding solution in Newtonian gravity, indicating the
possibility of describing the relativistic gravitational field by means of
multipole moments. In 1918, Lense and Thirring \cite{A3} discovered
an approximate exterior solution which, apart from the mass
monopole, contains an additional parameter that can be interpreted
as representing the angular momentum of the massive body. From this
solution it became clear that in Einstein's relativistic theory
rotation generates a gravitational field that leads to the dragging
of inertial frames (Lense-Thirring effect). This is the so--called
gravitomagnetic field which is of especial importance in the case of
rapidly rotating compact objects. The case of a static axisymmetric
solution with monopole and quadrupole moment was analyzed in 1959 by
Erez and Rosen \cite{erro59} by using spheroidal coordinates which are
specially adapted to describe the gravitational field of
non-spherically symmetric bodies. The exact exterior solution which
considers arbitrary values for the angular momentum was found by
Kerr \cite{A10} only in 1963. The problem of finding exact solutions changed
dramatically after Ernst \cite{ernst} discovered in 1968 a new
representation of the field equations for stationary axisymmetric
vacuum solutions. In fact, this new representation was the starting
point to investigate the Lie symmetries of the field equations.
Today, it is known that for this special case the field equations
are completely integrable and solutions can be obtained by using the
modern solution generating techniques \cite{dietz}. 
In this work, we
will analyze a particular class of solutions, derived by Quevedo and
Mashhoon \cite{A15}
 in 1991, which in the most general case contains infinite sets of gravitational and electromagnetic multipole moments.

As for the interior gravitational field of compact objects, the
situation is more complicated. There exists in the literature a
reasonable number of interior spherically symmetric solutions
\cite{intsch} which can be matched with the exterior Schwarzschild
metric. Nevertheless, a major problem of classical general
relativity consists in finding a physically reasonable interior
solution for the exterior Kerr metric. Although it is possible to
match numerically the Kerr solution with the interior field of an
infinitely tiny rotating disk of dust \cite{meinel}, such a
hypothetical system does not seem to be of relevance to describe
astrophysical compact objects. It is now widely believed that the
Kerr solution is not appropriate to describe the exterior field of
rapidly rotating compact objects. Indeed, the Kerr metric takes into
account the total mass and the angular momentum of the body.
However, the moment of inertia is an additional characteristic of
any realistic body which should be considered in order to correctly
describe the gravitational field. As a consequence, the multipole
moments of the field created by a rapidly rotating compact object
are different from the multipole moments of the Kerr metric
\cite{stergioulas}. For this reason a solution with arbitrary sets
of multipole moments, such as the one presented in this work, can be used to
describe the exterior field of arbitrarily rotating mass
distributions.

\section{Line element and field equations} 
\label{sec:line}

Although there exist in the literature many suitable coordinate systems, 
stationary axisymmetric gravitational fields are usually described
in cylindric coordinates $(t,\rho,z,\varphi)$. Stationarity implies that
$t$ can be chosen as the time coordinate and the metric does not depend
on time, i.e. $\partial g_{\alpha\beta}/\partial t =0$. Consequently, the corresponding 
timelike Killing vector has the components $\delta^\alpha_t$. A second Killing
vector field is associated to the axial symmetry with respect to the axis 
$\rho=0$.  Then, choosing $\varphi$ as the azimuthal angle, the metric satisfies
the conditions $\partial g_{\alpha\beta}/\partial \varphi =0$, and the components of 
the corresponding spacelike Killing vector are $\delta^\alpha_\varphi$.

Using further the properties of stationarity and axial symmetry, together
with the vacuum field equations, for a general metric of the form 
$g_{\alpha\beta}=g_{\alpha\beta}(\rho,z)$, it is possible to show that the most general 
line element for this type of gravitational fields can be written
in the Weyl-Lewis-Papapetrou form as \cite{weyl,lewis,pap,solutions}
\be
ds^2 = f(dt-\omega d\varphi)^2 - 
f^{-1}\left[e^{2 k  }(d\rho^2+dz^2) +\rho^2d \varphi^2\right] \ ,
\label{lel}
\ee
where $f$, $\omega$ and $k$ are functions of $\rho$ and $z$, only. 
After some rearrangements which include the introduction of a new function 
$\Omega=\Omega(\rho,z)$ by means of
\be
\rho \partial_\rho \Omega = f^2 \partial_z \omega \ , 
\qquad \rho \partial_z\Omega = - f^2   \partial_\rho \omega \ ,
\label{bigomega}
\ee 
the vacuum field equations $R_{\alpha\beta}=0$ can be
shown to be equivalent to the following set of partial differential equations
\be
\frac{1}{\rho}\partial_\rho(\rho\partial_\rho f)  + \partial_z^2 f + \frac{1}{f}[
(\partial_\rho \Omega) ^2  + (\partial_z \Omega) ^2 - (\partial_\rho f) ^ 2 -  
 (\partial_z f) ^2]
=0 \ ,
\label{main1}
\ee
\be
\frac{1}{\rho}\partial_\rho(\rho\partial_\rho \Omega)  + \partial_z^2 \Omega 
-  \frac{2}{f}\left(
\partial_\rho f\, \partial_\rho\Omega + \partial_z f \,\partial_z\Omega\right)
=0 \ ,
\label{main2}
\ee
\be
\partial_\rho k = \frac{\rho}{4f^2}\left[ (\partial_\rho f)^2+ (\partial_\rho \Omega)^2
 - (\partial_z f)^2  - (\partial_z \Omega)^2\right] \ ,
\label{krho}
\ee
\be
\partial_z k = \frac{\rho}{2f^2}\left( \partial_\rho f \ \partial_z f 
+ \partial_\rho \Omega \ \partial_z \Omega \right)\ .
\label{kz}
\ee
It is clear that the field equations for $k$ can be integrated by quadratures, 
once $f$ and $\Omega$ are known. For this reason, the equations 
(\ref{main1}) and (\ref{main2}) for $f$ and $\Omega$ are usually considered 
as the main field equations for stationary axisymmetric vacuum 
gravitational fields. 

It is interesting to mention that the main field equations can be obtained from a Lagrangian in the
following way. The Einstein-Hilbert Lagrangian ${\cal L}_{EH}=\sqrt{-g}R$ for the line element (\ref{lel}) with 
the auxiliary function $\Omega$, as defined in Eq.(\ref{bigomega}), can be written as 
\be 
{\cal L}_{EH} =
 \frac{\rho}{2f^2}\left[
(\partial_\rho f)^2 +  (\partial_z f)^2 +
 (\partial_\rho \Omega)^2 +  (\partial_z \Omega)^2\right]  \ ,
\label{lsig} 
\ee
where the terms containing second order derivatives have been eliminated by neglecting the total divergence terms, and
a Legendre transformation has been applied for the ``cyclic'' functions $\Omega$ and $k$ \cite{cnq01}. Then, the variation
of this Lagrangian density with respect to $f$ and $\Omega$ generates the main field equations (\ref{main1}) and (\ref{main2}).

%%%%%%%%%%%%%%%%%%%%%%%%%%%%%%%%%%%%%%%%%%%%%%%%%%%%%%%%%%%%%%%%
%%%%%%%%%%%%%%%%%%%%%%%%%%%%%%%%%%%%%%%%%%%%%%%%%%%%%%%%%%%%%%% 
\subsection{Generalized harmonic maps}
\label{sec:examples}

An alternative differential geometric interpretation of stationary axisymmetric
gravitational fields can be explored by using the concepts of harmonic maps \cite{misner78} as follows.

Consider two (pseudo-)Riemannian manifolds $(M,\gamma)$ and $(N,G)$ 
of dimension $m$ and $n$, respectively.
Let $x^a$ and $X^\mu$ be coordinates on $M$ and $N$, respectively. 
This coordinatization implies that in general the 
metrics $\gamma$ and $G$ become functions of the corresponding 
coordinates. Let us assume that not only $\gamma$ 
but also $G$ can explicitly depend on the coordinates $x^a$, i.e.
let $\gamma=\gamma(x)$ and $G=G(X,x)$. 
A smooth map $X: M\rightarrow N$ will be called an 
$(m\rightarrow n)-$generalized harmonic
map if it satisfies the Euler-Lagrange equations 
\be
\frac{1}{\sqrt{|\gamma|}}\partial_b\left(\sqrt{|\gamma|}\gamma^{ab}
  \partial_a X^\mu \right) + \Gamma^\mu_{\ \nu\lambda } \, \gamma^{ab} \,
  \partial_a X^\nu  \partial_b X^\lambda + G^{\mu\lambda} \gamma^{ab} \, 
  \partial_a X^\nu \, \partial_b G_{\lambda\nu} 
 = 0 \ ,
\label{gengeo}
\ee
which follow from the variation of the generalized action  
\be  
\label{genact}
S = \int d^m x \sqrt{|\gamma|}\, \gamma^{ab}(x)\, \partial_a \, X^\mu
\partial_ b X^\nu  G_{\mu\nu}(X,x)   \ ,
\ee
with respect to the fields $X^\mu$. Here the Christoffel symbols, 
determined by the metric $G_{\mu\nu}$, are calculated in the standard 
manner, without considering the explicit dependence on $x$.  
Notice that the presence of the
term $G_{\mu\nu}(X,x)$ in the Lagrangian density takes into account the ``interaction''
 between the base space $M$ and the target
space $N$. This interaction leads to an extra term in the motion equations, 
as can be seen in (\ref{gengeo}), which is important in order to recover the 
correct field equations in the case of gravitational fields. Moreover,
this interaction affects the conservation laws 
of the physical systems we attempt to describe by means of generalized
harmonic maps. 
To see this explicitly we calculate the covariant derivative 
of the generalized 
Lagrangian
density 
\be
{\cal L} = \sqrt{|\gamma|}\, \gamma^{ab}(x)\, \partial_a \, X^\mu
\partial_ b X^\nu  G_{\mu\nu}(X,x) \ ,
\label{genlag0}
\ee
and replace the result in the corresponding
motion equations (\ref{gengeo}). 
Then, the final result can be written as
\be
\nabla_b  T _a^{\ b} + \frac{1}{2} \frac{\partial {\cal L}}{\partial x^a} = 0 \ ,
\label{claw2}   
\ee
where $T _{ab}$ represents the canonical energy-momentum tensor 
\be 
T _{ab} = \frac{\delta {\cal L}}{\delta \gamma^{ab}} \ ,\ \ T_a^{\ b}=
 \sqrt{\gamma} G_{\mu\nu} \left( 
\gamma^{bc} \partial_a X^\mu \, \partial_c X^\nu - 
\frac{1}{2}\delta_a^b \gamma^{cd} \partial_c X^\mu \, \partial_d X^\nu\right).  
\label{emt}
\ee 
The standard conservation law $(\nabla_b  T _a^{\ b} =0)$ is recovered only when the Lagrangian does not
depend explicitly on the coordinates of the base space. Even if we choose a 
flat base space $\gamma_{ab} = \eta_{ab}$, the explicit dependence of the metric
of the target space $G_{\mu\nu}(X,x)$ on $x$ generates a term that violates the
standard conservation law. This term is due to the interaction between 
the base space and the target space which, consequently, 
is one of the main characteristics of
the generalized harmonic maps.

Consider a $(2 \to 2)-$generalized harmonic map. 
Let $x^a=(\rho,z)$ be the coordinates on the base space $M$,
and $X^\mu=(f,\Omega)$ the coordinates on the target space $N$.
In the base space we choose a flat metric and  in the target space 
a conformally flat metric, i.e. 
 \begin{equation} 
 \label{explmet}
\gamma_{ab} = \delta_{ab} \qquad {\rm and} \qquad G_{\mu\nu}=\frac{\rho}{2f^2}\delta_{\mu\nu} \qquad (a,b=1,2; \ \mu,\nu=1,2).
\end{equation} 
A straightforward computation shows that the generalized Lagrangian (\ref{genlag0}) 
coincides with the Lagrangian (\ref{lsig}) for stationary axisymetric fields, and 
that the equations of motion (\ref{gengeo}) generate the main field equations
(\ref{main1}) and (\ref{main2}).  Moreover, if we calculate  
the components of the energy-momentum 
tensor $ T _{ab}=\delta {\cal L}/\delta \gamma^{ab}$, we obtain  
\begin{equation}   \label{emtaxi2}
T_{\rho\rho} = -T_{zz} =  \frac{\rho}{4f^2}\left[ (\partial_\rho f)^2 + (\partial_\rho \Omega)^2 - (\partial_z f)^2  - (\partial_z \Omega)^2 \right],
\end{equation}
\begin{equation}  
 \label{emtaxi3}
T_{\rho z} = \frac{\rho}{2f^2}\left( \partial_\rho f \,\partial_z f + \partial_\rho \Omega \,\,\partial_z \Omega \right).
\end{equation}
This tensor is traceless due to the fact that the base space is 2-dimensional. 
It satisfies the generalized conservation law (\ref{claw2})
on-shell:
\be 
\frac{d T_{\rho\rho}}{d \rho} + \frac{d T_{\rho z}}{d z} +\frac{1}{2}
\frac{\partial {\cal L}}{\partial \rho} = 0 \ ,
\label{claw3}
\ee
\be
\frac{d T_{\rho z}}{d \rho} - \frac{d T_{\rho \rho}}{d z} = 0 \ . 
\label{claw4}
\ee
Incidentally, the last equation coincides with the integrability condition for the metric
function $k$, which is identically satisfied by virtue of the main field equations. 
In fact, as can be seen from Eqs.(\ref{krho},\ref{kz}) and 
(\ref{emtaxi2},\ref{emtaxi3}), the components of the energy-momentum tensor 
satisfy the relationships 
$T_{\rho\rho} = \partial_\rho k$ and $ T_{\rho z} = \partial_z k$, so that 
the conservation law (\ref{claw4}) becomes an identity. Although we have eliminated
from the starting Lagrangian (\ref{lsig}) the variable $k$ by applying a Legendre 
transformation on the Einstein-Hilbert Lagrangian (see \cite{cnq01} for details)
for this type of gravitational fields, the formalism of generalized harmonic maps
seems to retain the information about $k$ at the level of the generalized 
conservation law.

The above results show that stationary axisymmetric spacetimes
can be represented as a $(2\to 2)-$generalized harmonic map with metrics given as
in (\ref{explmet}). 
It is also possible to interpret the generalized harmonic map given above 
as a generalized string model. Although the metric of the base space $M$ is Euclidean,
we can apply a Wick rotation $\tau=i\rho$ to obtain a Minkowski-like structure on $M$. 
Then, $M$ represents the world-sheet of a bosonic string in which $\tau$ is measures 
the time and $z$ is the parameter along the string. The string is ``embedded'' in the
target space $N$ whose metric is conformally flat and explicitly depends on the 
time parameter $\tau$. For more details see \cite{hnq09}.

%%%%%%%%%%%%%%%%%%%%%%%%%%%%%%%%%%%%%%%%%%%%%%%%%%%%%%%%%%%%%%%%%%%%%
%%%%%%%%%%%%%%%%%%%%%%%%%%%%%%%%%%%%%%%%%%%%%%%%%%%%%%%%%%%%%%%%%%%%%
\section{The static solution}

Let us consider the special case of static axisymmetric fields. This corresponds
to metrics which, apart from being axially symmetric and independent of the time 
coordinate, are invariant with respect to the transformation $\varphi \rightarrow 
-\varphi$ (i.e. rotations with respect to the axis of symmetry are not allowed). 
Consequently, the corresponding line element is given by 
\be
ds^2 = f dt^2 - 
f^{-1}\left[e^{2 k  }(d\rho^2+dz^2) +\rho^2d \varphi^2\right] \ ,
\label{lel1}
\ee
and the field equations can be written as 
\be
\partial_\rho^2 \psi + \frac{1}{\rho}\partial_\rho \psi + \partial_z^2 \psi = 0 \ ,
\quad f=\exp(2\psi)\ ,\
\label{eqpsi}
\ee
\be
\partial_\rho k =  \rho\left[ (\partial_\rho \psi)^2 - (\partial_z \psi)^2\right]\ ,
\quad 
\partial_z k = 2 \rho  \partial_\rho \psi \ \partial_z \psi  \ . 
\label{eqkstatic}
\ee
We see that the main field equation (\ref{eqpsi}) corresponds to the linear 
Laplace equation for the metric function $\psi$. 
The general solution of 
Laplace's equation is known and, if we demand additionally asymptotic flatness, 
we obtain the Weyl solution which can be written as \cite{weyl,solutions} 
\be
\psi = \sum_{n=0}^\infty \frac{a_n}{(\rho^2+z^2)^\frac{n+1}{2}} P_n({\cos\theta}) \ ,
\qquad \cos\theta = \frac{z}{\sqrt{\rho^2+z^2}} \ ,
\label{weylsol}
\ee
where $a_n$ $(n=0,1,...)$ are arbitrary constants, and $P_n(\cos\theta)$ represents the Legendre
polynomials of degree $n$. 
The expression for the metric function $\gamma$ 
can be calculated by quadratures by using the set of first order differential 
equations (\ref{eqkstatic}). Then 
\be
\gamma = - \sum_{n,m=0}^\infty \frac{ a_na_m (n+1)(m+1)}{(n+m+2)(\rho^2+z^2)^\frac{n+m+2}{2} }
\left(P_nP_m - P_{n+1}P_{m+1} \right) \ .
\ee
Since this is the most general static, axisymmetric, asymptotically flat
vacuum solution, it must contain all known solutions of this class. In particular,
one of the most interesting special solutions which is Schwarzschild's spherically symmetric 
black hole spacetime must be contained in this class. To see this, we must choose
the constants $a_n$ in such a way that the infinite sum (\ref{weylsol}) converges to the
Schwarzschild solution in cylindric coordinates. But, of course, this representation 
is not the most appropriate to analyze the interesting physical properties of 
Schwarzschild's metric. 

In fact, it turns out that  to investigate the properties 
of solutions with multipole moments it is more convenient to use prolate spheroidal coordinates
$(t,x,y,\varphi)$ in which the line element can be written as
\be
 ds^2 = f dt^2 - \frac{\sigma^2}{f}\left[ e^{2k}(x^2-y^2)\left( \frac{dx^2}{x^2-1} + \frac{dy^2}{1-y^2} \right) 
+ (x^2-1)(1-y^2) d\varphi^2\right] 
\label{lelxy}
\ee
where
\be 
x= \frac{r_++r_-}{2\sigma} \ ,\quad (x^2\geq 1),\quad  y = \frac{r_+-r_-}{2\sigma} \ , \quad (y^2 \leq 1) 
\ee
\be 
r_\pm^2 = \rho^2 + (z\pm \sigma)^2\ ,\quad  
\sigma=const\ ,
\ee 
and the metric functions  $f$, and $k$ depend on $x$ and $y$, only.
In this coordinate system, the main field equation becomes
\be
[(x^2-1)\psi_x]_x + [(1-y^2)\psi_y]_y =0\ ,\quad f=\exp(2\psi) \ ,
\label{eqpsixy}
\ee
and the general static solution which is also asymptotically flat 
can be expressed as
\be
 \psi = \sum_{n=0}^\infty (-1)^{n+1} q_n P_n(y) Q_n (x) \ , \quad q_n = const 
\label{gensolxy}
\ee
where $P_n(y)$ are the Legendre polynomials, and $Q_n(x)$ are the Legendre functions of second kind. In particular,
\be
P_0=1, \quad P_1 = y ,\quad P_2 = \frac{1}{2} (3y^2-1) \ , ...
\nonumber
\ee
\be
Q_0 = \frac{1}{2}\ln\frac{x+1}{x-1}\ , \quad
Q_1 =  \frac{1}{2}x\ln\frac{x+1}{x-1}-1\ , \quad
\nonumber
\ee
\be
\nonumber
Q_2 = \frac{1}{2}(3x^2-1) \ln\frac{x+1}{x-1}-\frac{3}{2} x \ , ...
\ee
The corresponding function $k$ can be calculated by quadratures and its general expression has been explicitly 
derived in \cite{quev89}. 

The most important special cases contained in this general solution are the Schwarzschild 
metric 
 \be
\label{schxy}
\psi = - q_0 P_0(y)Q_0(x)= \frac{1}{2}\ln \frac{x-1}{x+1} \ , \quad k = \frac{1}{2}\ln \frac{x^2-1}{x^2-y^2}\ .
\ee
Indeed, the coordinate transformation 
\be
y = \cos\theta\ , \quad x = \frac{r-m}{\sigma}
\label{trart}
\ee
transforms the line element (\ref{lelxy}) into
\bea
ds^2 = & f dt^2 - \frac{1}{f}\bigg[ e^{2k}\left( 1-\frac{2m}{r} + \frac{m^2-\sigma^2\cos^2\theta}{r^2}\right) 
\left(\frac{dr^2}{1-\frac{2m}{r} + \frac{m^2-\sigma^2}{r^2}}+ r^2d\theta^2\right) \nonumber\\
& + r^2\left(1-\frac{2m}{r} 
+ \frac{m^2-\sigma^2}{r^2}\right) r^2\sin^2\theta d\varphi^2 \bigg]\ ,
\label{lelrt}
\eea
and the metric functions (\ref{schxy}) with $q_0=1$ into 
\be
\psi = \frac{1}{2}\ln\left(1- \frac{2m}{r}\right) \ ,\quad
k = \frac{1}{2} \ln \left(\frac{1-\frac{2m}{r} + \frac{m^2-\sigma^2}{r^2}}{1-\frac{2m}{r} + \frac{m^2-\sigma^2\cos^2\theta}{r^2}}\right) \ ,
\ee
which, when inserted in the above line element with $\sigma=m$, lead to the standard Schwarzschild solution in spherical coordinates
\be 
ds^2= \left(1-\frac{2m}{r}\right)dt^2 - \frac{dr^2}{1-\frac{2m}{r}} - r^2(d\theta^2+\sin^2\theta d\varphi^2)\ .
\ee
Moreover, the Erez-Rosen metric \cite{erro59} is obtained in the special case
\bea
\psi & = & -  P_0(y)Q_0(x) - q_2 P_2(y)Q_2(x) \nonumber \\
 & = & \frac{1}{2}\ln\left(\frac{x-1}{x+1}\right) + \frac{1}{2}q_2 (3y^2-1) \left[\frac{1}{4}(3x^2-1)\ln\left(\frac{x-1}{x+1}\right)+\frac{3}{2} x\right] \,
\eea
and
\bea
k = & &\frac{1}{2}(1+q_2)^2 \ln\left( \frac{x^2-1}{x^2-y^2}\right) - \frac{3}{2} q_2(1-y^2)\left[x \ln\left(\frac{x-1}{x+1}\right) + 2\right]\nonumber\\
& + & \frac{9}{16}q_2^2(1-y^2)\bigg[ x^2+4y^2 - 9 x^2y^2-\frac{4}{3} 
+ x\left(x^2+7y^2-9x^2y^2-\frac{5}{3}\right)\ln\left(\frac{x-1}{x+1}\right) \nonumber\\
& + & \frac{1}{4}(x^2-1)(x^2+y^2-9x^2y^2-1)\ln^2\left(\frac{x-1}{x+1}\right)\bigg]\ .
\eea
In the last case, the constant parameter $q_2$ turns out to determine the quadrupole moment. In general, the 
constants $q_n$ represent an infinite set of parameters that determines an infinite set of mass multipole moments. In fact, using 
the Geroch--Hansen \cite{ger,hans} definition, one can prove that the relativistic multipole moments
can be expressed as 
\be 
M_n = N_n + R_n \ ,\quad N_n = (-1)^n \frac{n!}{(2n+1)!!} q_n \sigma^{n+1} \ , \quad n=0,1,2,..., 
\ee
where $N_n$ are the Newtonian multipole moments which have been calculated by using the coordinate 
invariant method proposed by Ehlers in \cite{ehlers}. Moreover, the second term $R_n$ represents the relativistic corrections
\be
R_0=R_1=R_2=0\ ,\quad R_3 = -\frac{2}{5}\sigma^2 N_1 \ , \quad R_4 = - \frac{2}{7} \sigma^2 N_2 - \frac{6}{7} \sigma N_1^2 \ , \ ....
\ee
which in general can be determined in terms of lower Newtonian moments, i. e., $R_n= R_n (N_{n-2},N_{n-3},..., N_0)$.

The metric function given in Eq.(\ref{gensolxy}), together with the corresponding function $k$ derived in \cite{quev89}, 
represents the most general static axisymmetric solution of vacuum Einstein's equations in prolate spheroidal coordinates. 
Its multipolar structure represented by the infinite set of parameters $q_n$ can be used to describe the exterior 
gravitational field of any static mass distribution that preserves the axial symmetry. 

The simplest metric contained 
in this class of solutions corresponds to the Schwarzschild spacetime which describes the exterior gravitational field
of a black hole of mass $m$. According to the black hole uniqueness theorems the Schwarzschild spacetime represents the only 
static black hole, i. e., it is the only solution with a singularity covered by an event horizon. It then follows that 
all the multipolar solutions with multipoles higher than the monopole one must be characterized by the presence of naked 
singularities. This has been shown explicitly for the Erez-Rosen solution in \cite{quev90} by using a numerical approach 
due to the complexity of the resulting curvature invariants. To be able to perform an analytical investigation it is necessary
to consider a simpler metric. To this end, we use the following property of the field equations for static fields. 
If $\psi_0$ and $k_0$ represent an exact static solution of the field equations (\ref{eqpsi}) and (\ref{eqkstatic}), then 
the functions $\delta\psi_0$ and $\delta^2 k_0$, with $\delta=$ const are also solutions to the same field equations. This property
was first discovered by Zipoy \cite{zip66} and Voorhees \cite{voor70}. Consider then the solution 
\be
\psi = \frac{\delta}{2}\ln\frac{x-1}{x+1}\ , \quad 
\quad k = \frac{\delta^2}{2}\ln \frac{x^2-1}{x^2-y^2}\ ,
\ee
which represents the simplest generalization of the Schwarzschild metric (\ref{schxy}) with $q_0=1$. Then, introducing spherical
coordinates by means of the relations (\ref{trart}), and choosing the parameters $\sigma=m$ and $\delta = 1-q$, the resulting 
solution can be written as
\bea
ds^2 = & & \left(1-\frac{2m}{r}\right)^{1-q} dt^2   \\
& -& \left(1-\frac{2m}{r}\right)^{q}\left[ \left(1+\frac{m^2\sin^2\theta}{r^2-2mr}\right)^{q(2-q)} \left(\frac{dr^2}{1-\frac{2m}{r}}+ r^2d\theta^2\right) + r^2 \sin^2\theta d\varphi^2\right]\nonumber .
\label{zv}
\eea

This solution is axially symmetric and reduces to the spherically symmetric Schwarzschild metric in the limit $q\rightarrow 0$. 
It is asymptotically flat for any finite values of the parameters $m$ and $q$. 
To find the physical meaning of these parameters, 
we calculate the multipole moments of the solution by using the invariant definition proposed by Geroch \cite{ger}. The 
lowest mass multipole moments $M_n$, $n=0,1,\ldots $ are given by
\be 
M_0= (1-q)m\ , \quad M_2 = \frac{m^3}{3}q(1-q)(2-q)\ ,
\ee
whereas higher moments are proportional to $mq$ and can be 
completely rewritten in terms of $M_0$ and $M_2$. 
This means that the arbitrary parameters $m$ and $q$ determine the mass and quadrupole 
which are the only independent multipole moments of the solution. In the limiting case $q=0$ only the monopole $M_0=m$ 
survives, as in the Schwarzschild spacetime. In the limit $m=0$, with $q\neq 0$, all moments vanish identically, implying that 
no mass distribution is present and the spacetime must be flat. This can be seen also at the level of the curvature which vanishes in 
the limiting case $m\rightarrow 0$. This means that, independently of the value of $q$, there exists a coordinate transformation that 
transforms the resulting metric into the Minkowski solution. From a physical point of view this is an important  
property because it means that the parameter $q$ is related to a genuine mass distribution, i.e., there is no quadrupole moment 
without mass.  Furthermore, notice that all odd multipole moments are zero because the solution possesses an additional 
reflection symmetry with respect to the equatorial plane. 

We conclude that the above metric describes the exterior gravitational 
field of a static deformed mass. The deformation is described by the quadrupole moment $M_2$ which is positive for a prolate mass 
distribution and negative for an oblate one. Notice that in order to avoid the appearance of
a negative total mass $M_0$ the condition $q<1$ must be satisfied .

\begin{figure}
\includegraphics[scale=0.2]{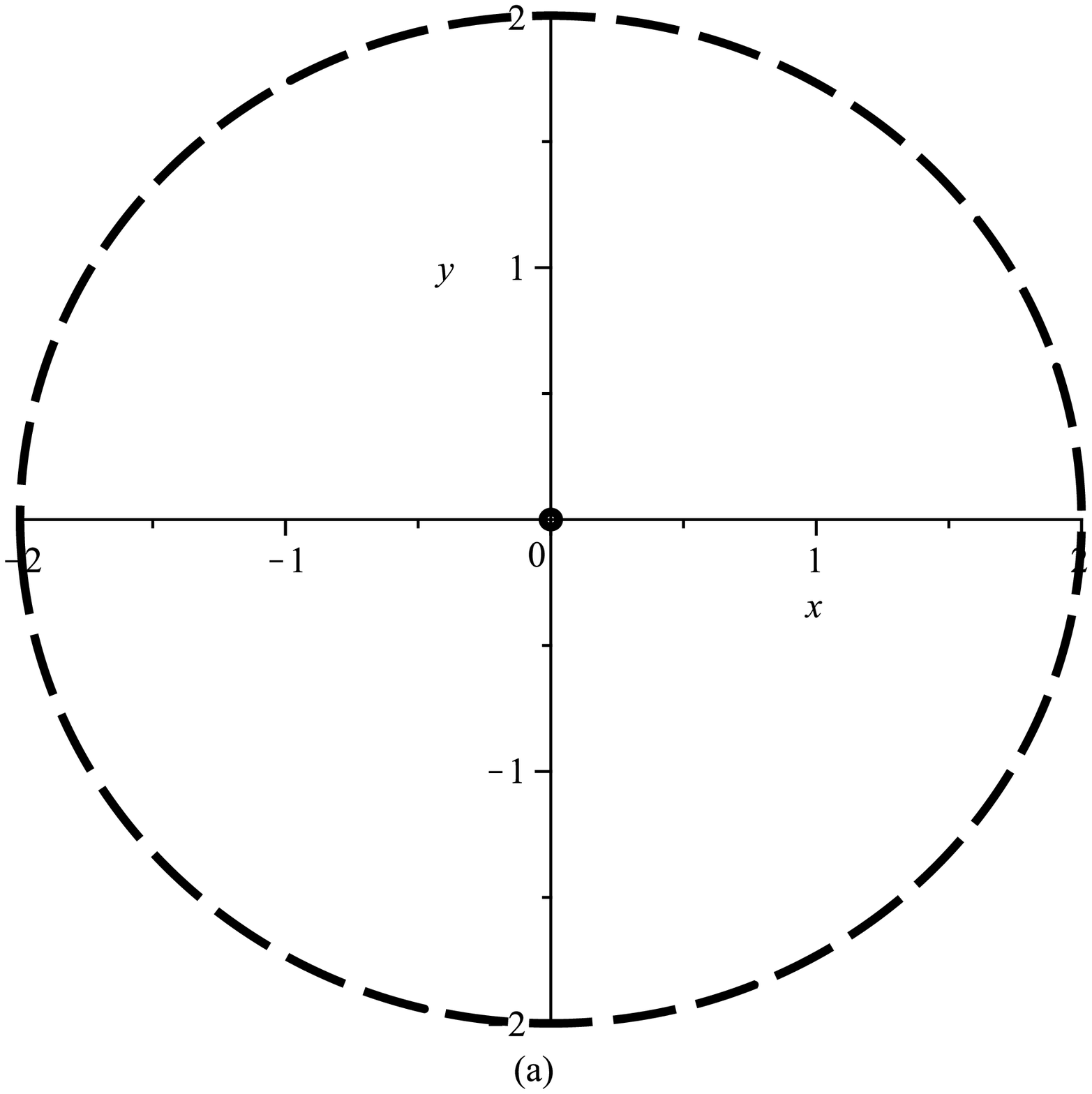}
\includegraphics[scale=0.2]{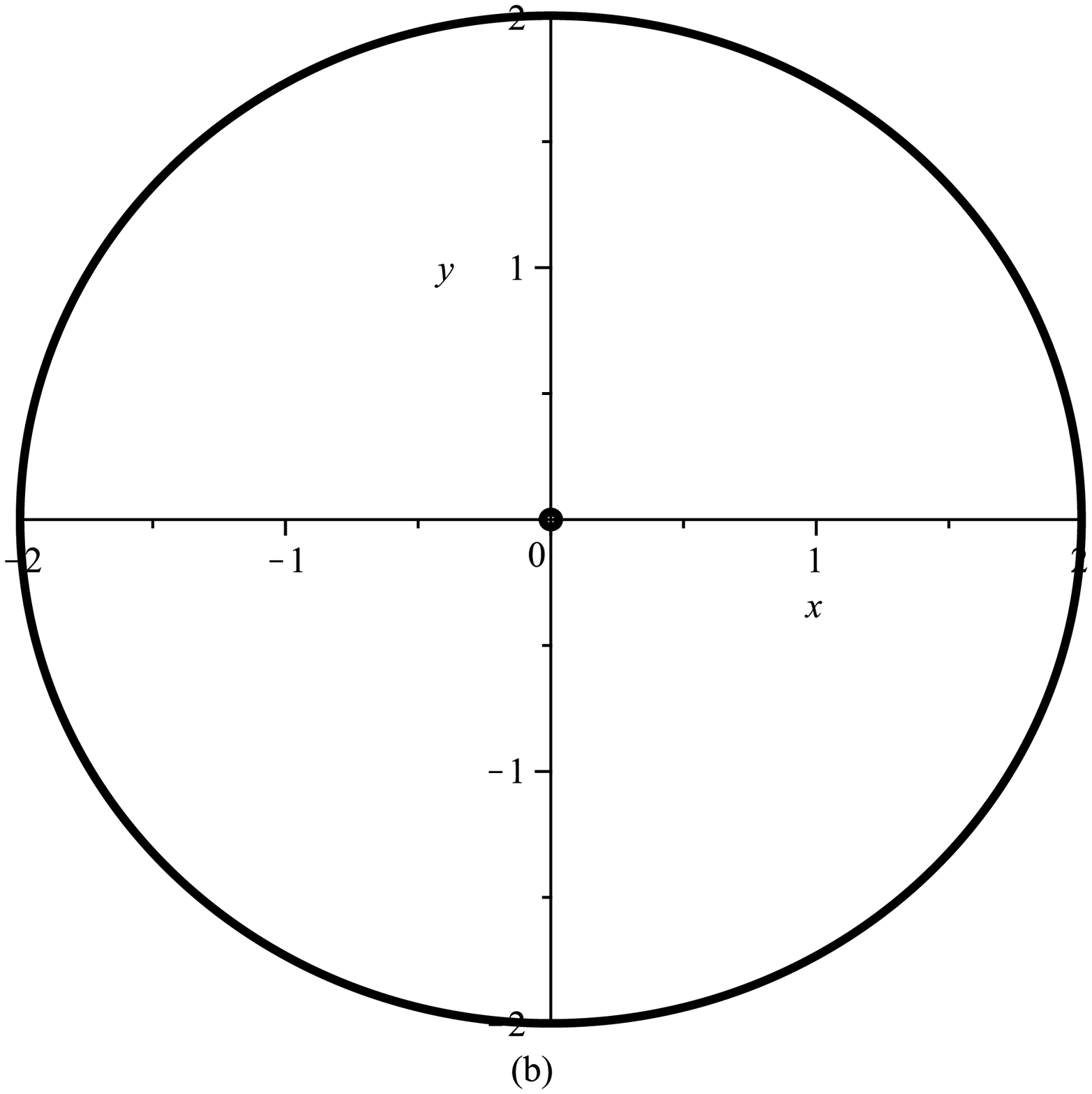}
\includegraphics[scale=0.2]{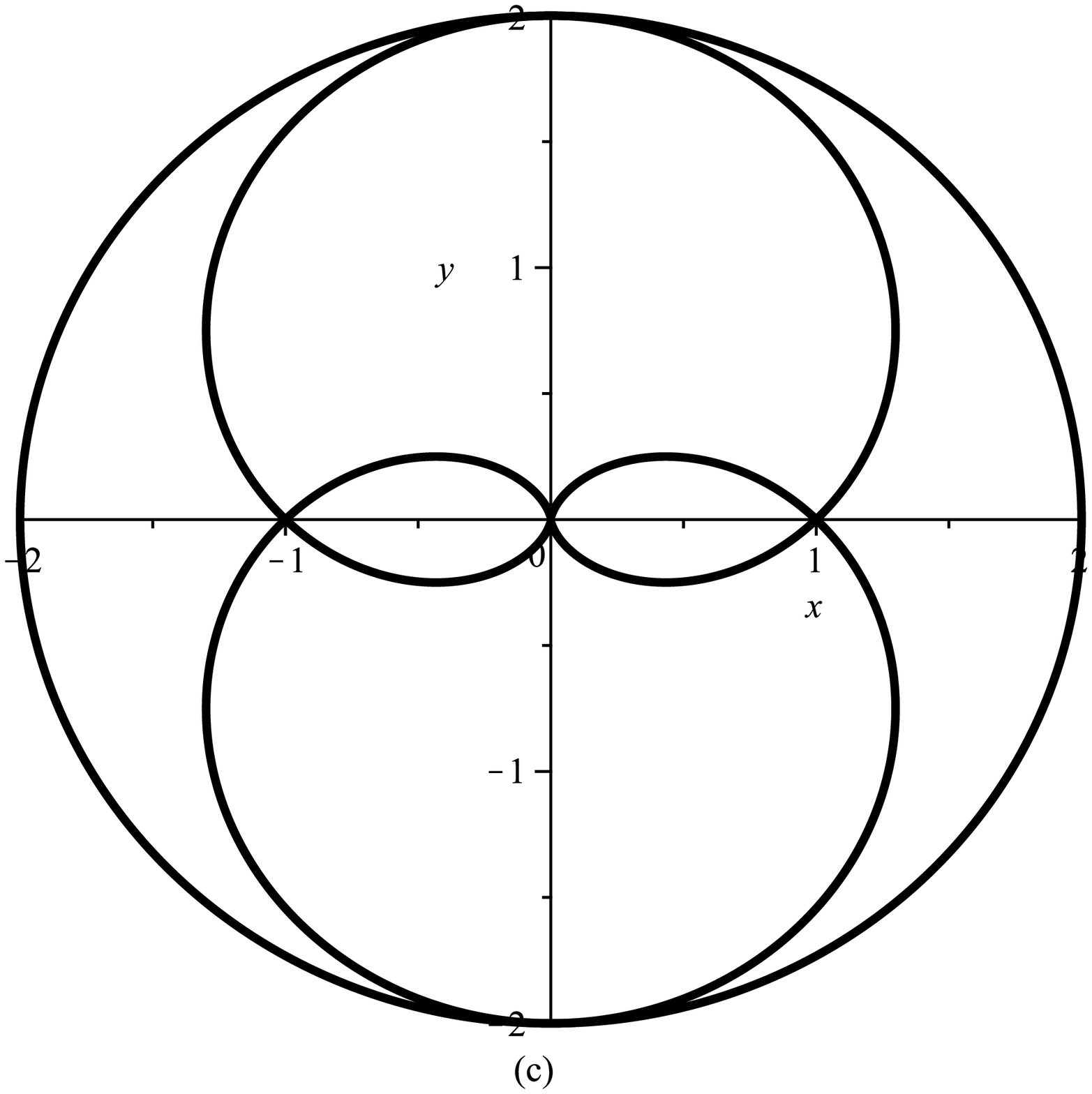}
\caption{Structure of naked singularities in a spacetime with quadrupole parameter $q$. 
For any value $q\neq 0$, there exists at least two naked singularities situated at $r=0$ and $r=2m$ as shown in 
plot (b) with solid curves. Furthermore, if the quadrupole parameter is contained within the interval 
$q\in \left(1- \sqrt{3/2},1+\sqrt{3/2}\right)\backslash \{0\}$, two additional
naked singularities appear as depicted in plot (c). The limiting case 
of a Schwarzschild spacetime $(q=0)$ with a singularity at the origin of coordinates surrounded by the horizon (dashed curve) 
situated at $r=2m$ is illustrated in plot (a).}
\label{fig1}
\end{figure}

To investigate the structure of possible curvature singularities, we consider the Kretschmann scalar 
$K = R_{\alpha\beta\gamma\delta}R^{\alpha\beta\gamma\delta}$. A straightforward computation leads to 
\be
K   = \frac{16 m^2(1-q)^2}{r^{4(2-2q+q^2)}}\frac{ (r^2-2mr+m^2\sin^2\theta)^{2q^2 -4q -1}}{(1-2m/r)^{2(q^2-q+1)}}L(r,\theta)\ ,
\label{kre}
\ee
with 
\bea
L(r,\theta)= & & 3(r-2m+qm)^2(r^2-2mr+m^2\sin^2\theta) \nonumber\\
& & - q(2-q)m^2\sin^2\theta[ q(2-q)m^2  + 3(r-m)(r-2m+qm)] \ .
\eea
In the limiting case $q=0$, we obtain the Schwarzschild value $K= {48 m^2}/{r^6}$ with the only singularity 
situated at the origin of coordinates $r\rightarrow 0$. In general, one can show that the singularity at
the origin, $r=0$, is present for any values of $q$. Moreover, an additional singularity appears at the radius $r=2m$ 
which, according to the metric (\ref{zv}), is also a horizon in the sense that the norm of the timelike Killing 
tensor vanishes at that radius. Outside the hypersurface $r=2m$ no additional horizon exists, indicating 
that the singularities situated at the origin and at $r=2m$ are naked. Moreover, for values of the quadrupole parameter 
within the interval
\be 
q\in \left(1- \sqrt{3/2},1+\sqrt{3/2}\right)\backslash \{0\}
\ee
a singular hypersurface appears at a distance 
\be
r_\pm = m(1\pm \cos\theta)
\ee
from the origin of coordinates. This type of singularity is always contained within the naked singularity situated at the radius
$r=2m$, and is related to a negative total mass $M_0$ for $q>1$. Nevertheless, in the interval 
$q\in (1- \sqrt{3/2},1] \backslash \{0\}$ the singularity is generated by a more realistic source with positive mass. 
This configuration of naked singularities is schematically illustrated 
in Fig. \ref{fig1}. 

\begin{figure}
\includegraphics[scale=0.3]{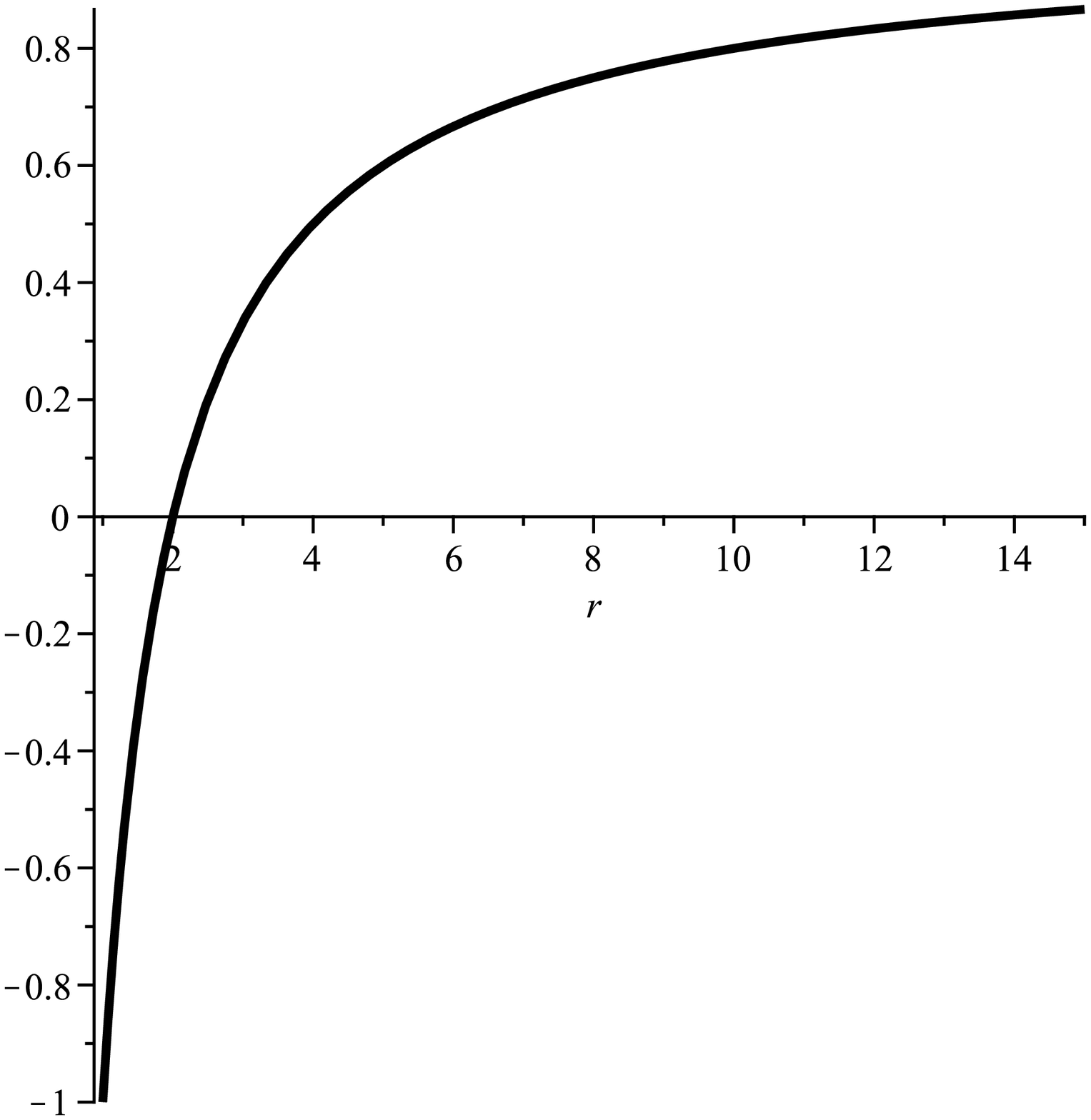}
\includegraphics[scale=0.3]{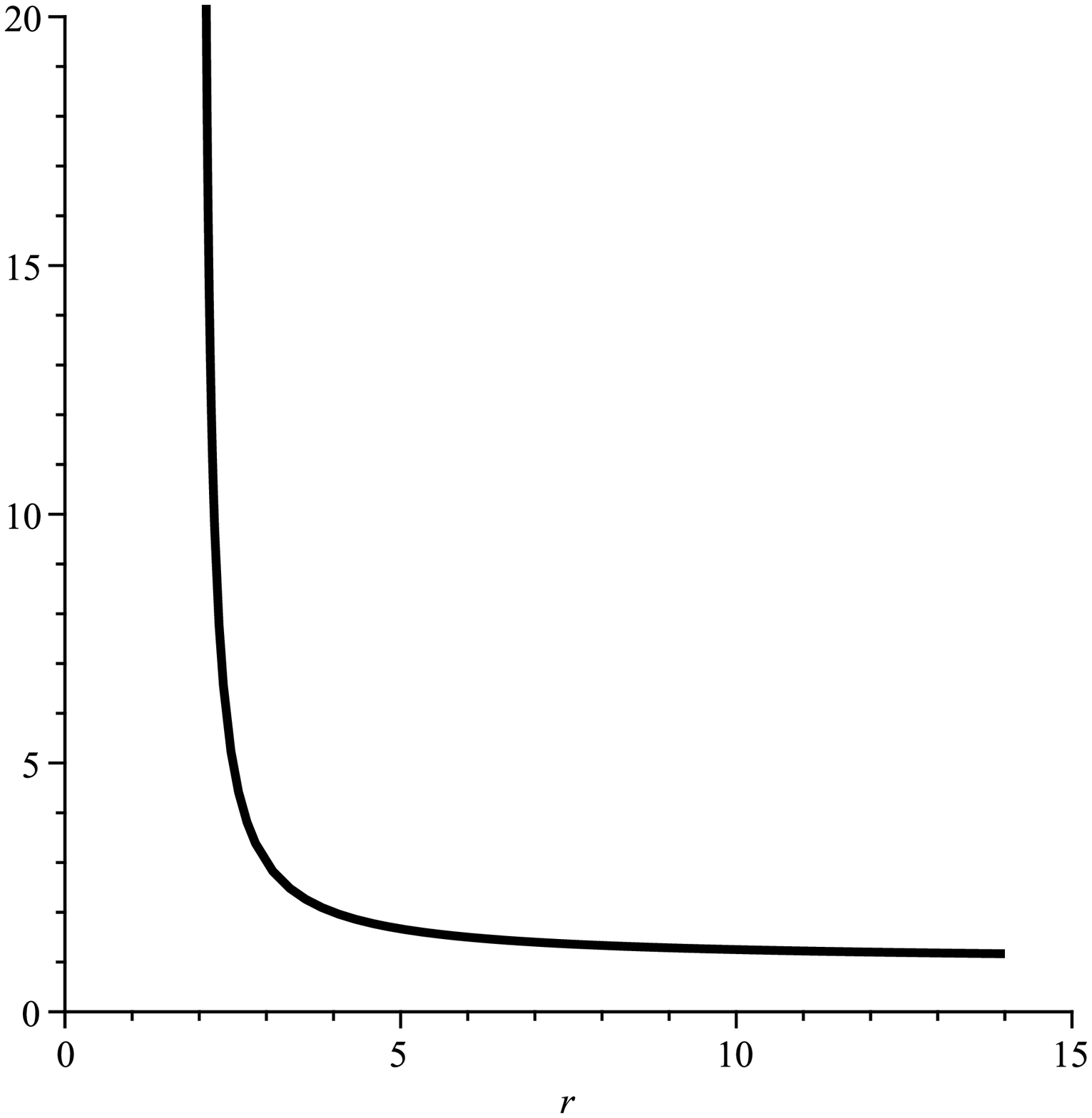}
\caption{The effective potential for the motion of timelike particles. Plot (a) shows the typical behavior of 
the effective potential  of a black hole configuration with $q=0$. The case of a naked singularity with $q=1/2$ 
is depicted in plot (b). } 
\label{fig2}
\end{figure}

Another important aspect related to the presence of naked singularities in multipolar solutions 
is the problem of repulsive gravity.  
In fact, it now seems to be established that naked singularities can appear as the result of 
a realistic gravitational collapse \cite{joshi07} and that naked singularities can 
generate repulsive gravity. Currently, there is no invariant definition of
repulsive gravity in the context of general relativity, although some attempts have been made by using 
invariant quantities constructed with the curvature of spacetime \cite{def89,def08,christian}. 
Nevertheless, it is possible
to consider an intuitive approach by using the fact that the motion of test particles in static axisymmetric
gravitational fields reduces to the motion in an effective potential. This is a consequence of the fact that
the geodesic equations possess two first integrals associated with stationarity and axial symmetry. The explicit 
form of the effective potential depends also on the type of motion under consideration. 
In the case of a massive test 
particle moving  along a geodesic contained in the equatorial plane  $(\theta = \pi/2)$ of the Zipoy--Voorhees 
spacetime (\ref{zv}), one can show that the effective potential reduces to
\be
V_{eff}^2 = \left(1-\frac{2m}{r}\right)^{1-q}\left[ 1 + \frac{L^2}{r^2}\left(1-\frac{2m}{r}\right)^{-q}\right]\ ,
\ee
where $L$ is constant associated to the angular momentum of the test particle as measured by a static observer at rest 
at infinity. This expression shows that the behavior of the effective potential strongly depends on the value of the quadrupole 
parameter $q$. This behavior is illustrated in Fig. \ref{fig2}. 

Whereas the effective potential of a black hole corresponds
to the typical potential of an attractive field, the effective potential of a naked singularity
is characterized by the presence of a barrier which acts on test particles as a source of repulsive gravity. 
We can show that in general in a mass distribution the presence of static multipoles higher than the monopole one leads to
the appearance of naked singularities in which effects associated with repulsive gravity can be found. 
It is therefore clear that the mass quadrupole and higher multipoles can be considered as sources 
of naked singularities in general relativity.
  
%%%%%%%%%%%%%%%%%%%%%%%%%%%%%%%%%%%%%%%%%%%%%%%%%%%%%%%%%%%%%%%%%%%%%%%%%%%
%%%%%%%%%%%%%%%%%%%%%%%%%%%%%%%%%%%%%%%%%%%%%%%%%%%%%%%%%%%%%%%%%%%%%%%%%%
\section{Stationary solutions}

The solution generating techniques \cite{dietz,solutions} can be applied, in particular, to any static seed solution 
in order to obtain the corresponding stationary generalization. One of the most powerful techniques is the 
inverse scattering  method  developed by Belinski and Zakharov \cite{bz78}. In this section we use a particular case of 
the inverse scattering method which is known as the Hoenselaers--Kinnersley-Xanthopoulos (HKX) 
transformation to derive the stationary  generalization of the general static solution in prolate spheroidal coordinates
presented in the last section.  

%%%%%%%%%%%%%%%%%%%%%%%%%%%%%%%%%%%%%%%%%
\subsection{Ernst representation}

In the general stationary case $(\omega\neq 0)$, the line element in prolate spheroidal coordinates  is given as 
\bea
& & ds^2 = f(dt-\omega d\varphi)^2 \nonumber \\
& - &
\frac{\sigma^2}{f}\left[ e^{2k}(x^2-y^2)\left( \frac{dx^2}{x^2-1} + \frac{dy^2}{1-y^2} \right) 
+ (x^2-1)(1-y^2) d\varphi^2\right] \ ,
\label{lelxystar}
\eea
where all the metric functions depend on $x$ and $y$ only. It turns out to be useful to introduce the 
the complex Ernst potentials
\be
E=f+i\Omega \ , \quad \xi = \frac{1-E}{1+E} \  ,
\ee
where the function $\Omega$ is now determined by the equations
\be
 \sigma (x^2-1)\Omega_x = f^2\omega_y\ ,\quad  \sigma (1-y^2)\Omega_y = - f^2 \omega_x \ . 
\ee
Then, it is easy to show that the main field equations can be represented in a compact and symmetric form  as 
\be 
(\xi \xi^*-1) \left\{[(x^2-1)\xi_x]_x + [(1-y^2)\xi_y]_y\right\} = 2 \xi^*[ (x^2-1)\xi_x^2 + (1-y^2)\xi_y^2]\ ,
\label{eqksixy}
\ee
where the asterisk represents complex conjugation. Notice that in the case of static fields the Ernst potentials 
become real and the above equation generates a linear differential equation for $\psi=(1/2)\ln f$. It is easy 
to see that the equation (\ref{eqksixy}) 
is invariant with respect to the transformation $x\leftrightarrow y$. Then, since the particular solution
\be
\xi = \frac{1}{x} \rightarrow \Omega=0 \rightarrow \omega = 0 \rightarrow k = \frac{1}{2}\ln \frac{x^2-1}{x^2-y^2}
\ee
represents the Schwarzschild spacetime, the choice $\xi^{-1} = y$ is also an exact solution. Furthermore, if we take the
linear combination $\xi^{-1} = c_1 x+c_2y$ and introduce it into the field equation (\ref{eqpsixy}), we obtain the new solution 
\be 
\xi^{-1} = \frac{\sigma}{m} x+i\frac{a}{m} y\ ,\  \sigma =\sqrt{m^2-a^2}\ ,
\ee
which corresponds to the Kerr metric in prolate spheroidal coordinates. The corresponding metric functions 
are 
\begin{eqnarray}
f&=&\frac{c^2x^2+d^2y^2-1}{(cx+1)^2+d^2y^2}\ , \quad 
\omega=2a\frac{(cx+1)(1-y^2)}{c^2x^2+d^2y^2-1}\ , \nonumber\\
k&=&\frac12\ln\left(\frac{c^2x^2+d^2y^2-1}{c^2(x^2-y^2)}\right)\ , 
\end{eqnarray}
where 
\be
c=\frac{\sigma}{m}\ , \quad d=\frac{a}{m}\ , \quad c^2+d^2=1\ .
\ee

In the case of the Einstein-Maxwell theory, the main field equations can be expressed as 
\be
(\xi\xi^* - {\cal F} {\cal F}^* -1)\nabla^2 \xi = 2(\xi^*\nabla\xi - {\cal F}^*\nabla {\cal F})\nabla \xi\ ,\ 
\ee
\be
(\xi\xi^* - {\cal F} {\cal F}^* -1)\nabla^2 {\cal F} = 2(\xi^*\nabla\xi - {\cal F}^*\nabla {\cal F})\nabla {\cal F}\ 
\ee
where $\nabla$ represents the gradient operator in prolate spheroidal coordinates. Moreover, 
 the gravitational potential $\xi$ and the electromagnetic ${\cal F}$ Ernst potential are defined as
\be
\xi = \frac{1-f-i\Omega}{1+f+i\Omega}\ , \quad {\cal F}=2\frac{\Phi}{1+f+i\Omega}\ .
\ee
The potential $\Phi$ can be shown to be determined uniquely by the electromagnetic potentials $A_t$ and $A_\varphi$ 
Furthermore, one can show that if $\xi_0$ is a vacuum solution, then the new potential
\be
\xi = \xi_0\sqrt{1-e^2}
\ee
represents
a solution of  the Einstein-Maxwell equations with effective electric charge $e$. 
This transformation is known in the literature as the Harrison transformation \cite{harrison}.
Accordingly, the Kerr--Newman 
solution in this representation acquires the simple form
\be
\xi = \frac{\sqrt{1-e^2}}{\frac{\sigma}{m} x+i\frac{a}{m}y} \ ,\qquad e = \frac{Q}{m}\ ,\qquad \sigma = \sqrt{m^2-a^2-Q^2} \ .
\ee
In this way, it is very easy to generalize any vacuum solution to include the case of electric charge. 
More general transformations of this type can be used in order to generate solutions with any desired set
of gravitational and electromagnetic multipole moments \cite{quev92}.

%%%%%%%%%%%%%%%%%%%%%%%%%%%%%%%%%%5
\section{The general solution}

If we take as seed metric the general static solution in prolate spheroidal coordinates with an arbitrary Zipoy--Voorhees 
parameter,
\be
\psi = \delta \sum_{n=0}^\infty (-1)^n q_n P_n(y)Q_n (x)\ ,
\ee
the application of two HXK transformations generates 
a stationary solution with an infinite number of gravitoelectric and gravitomagnetic multipole moments. 
The HKX method generates a new Ernst potential $\xi$ which can be written as 
\be 
\xi=  \frac{ (a_++ib_+)e^{2\delta\hat\psi} + a_-+i b_- } { (a_+ +ib_+)e^{2\delta\hat\psi} -a_--i b_-}(1-e_0^2 + g_0^2)^{1/2}\ ,
\quad {\Phi}= \frac{e_0+ig_0}{1+\xi}\ , 
\label{epot}
\ee
where
\be 
\hat\psi = \sum_{n=1}^\infty (-1)^n q_n P_n(y)Q_n(x)
\ee
\be
a_\pm = (x\pm1)^{\delta-1} [x(1-\lambda\mu) \pm (1+\lambda\mu)] \ , 
\ee
\be
b_\pm = (x\pm 1)^{\delta -1}[y(\lambda+\mu)\mp(\lambda-\mu)]\ ,
\ee
with
\be
\lambda = \alpha_1 (x^2-1)^{1-\delta}(x+y)^{2\delta -2} e^{2\delta \sum_{n=1}^\infty (-1)^n q_n\beta_{n-}},
\ee
\be
\mu = \alpha_2 (x^2-1)^{1-\delta}(x-y)^{2\delta -2} e^{2\delta \sum_{n=1}^\infty (-1)^n q_n\beta_{n+}}\ ,
\ee
and
\bea
\beta_{n\pm} = & & (\pm 1)^n \left[\frac{1}{2}\ln \frac{(x\mp y)^2}{x^2-1} - Q_1(x)\right] + P_n(y)Q_{n-1}(x)\nonumber \\
               & &-\sum_{k=1}^{n-1} (\pm1)^k P_{n-k}(y)\left[Q_{n-k+1}(x) - Q_{n-k-1}(x)\right]\ .
\eea
Here $P_n(y)$ and $Q_n(x)$ represent the Legendre polynomials and functions of second kind, respectively. The constant
parameters $e_0$, $g_0$, $\sigma$, $\alpha_1$, $\alpha_2$, $q_n$, and $\delta$ determine the gravitational and electromagnetic
multipole moments. The metric functions $f$ and $\omega$ can be obtained from the definitions of the Ernst potentials whereas
the function $k$ can be calculated by quadratures once $f$ and $\omega$ are known. 

In general, this solution is asymptotically flat and free of singularities along the axis of symmetry, $y=1$, outside 
certain region situated close to the origin of coordinates. The sets of infinite multipole moments can be chosen in such a
way as to reproduce the shape of ordinary axially symmetric compact objects. 
One of the most interesting solutions contained
in this family is the one with non-vanishing parameters $q_0=1$, $q_2=q$, $\delta$, $\alpha_1=\alpha_2= (\sigma - m)/a$, where 
$m$ and $a$ are new constants. In this case,  the solution possesses the following independent parameters: $m$, $a$, $\delta$, and $q$. 
In the limiting case $\alpha=0$, $a=0$, $q=0$ and $\delta =1$, the only independent parameter is $m$ and the Ernst potential (\ref{epot})
determines the Schwarzschild spacetime. Moreover, for $\alpha =a =0$ and $q=0$ we obtain the Ernst potential of the Zipoy-Voorhees (ZV)
\cite{zip66,voor70}
 static 
solution which is characterized by the parameters $m$ and $\delta$. Furthermore, for   $\alpha =a =0$ and 
$\delta =1$, the resulting solution coincides with the Erez-Rosen (ER) static spacetime \cite{erro59}. 
The Kerr metric is also contained as a special case for $q=0$ and $\delta =1$. 
The physical 
significance of the parameters entering this particular solution 
can be established in an invariant manner by calculating the relativistic Geroch--Hansen \cite{ger,hans} multipole moments.
We use here the procedure formulated in Ref. \cite{quev89} which allows us to derive the gravitoelectric $M_n$ as well as the 
gravitomagnetic $J_n$ multipole moments. A lengthy but straightforward calculation yields 
\be 
M_{2k+1} = J_{2k}=0 \ ,  \quad k = 0,1,2,... 
\ee 
\be 
M_0= m + \sigma(\delta -1)
\ee
\be
M_2 = \frac{2}{15} \sigma^3 \delta q - \frac{1}{3}\sigma^3 (\delta^3 -3\delta^2-4\delta + 6) - m\sigma^2 \delta (\delta -2) - 3m^2 \sigma (\delta -1)
-m^3 \ ,
\ee
\be
J_1 = ma + 2a \sigma (\delta -1)\ ,
\ee
\bea
J_3 && = \frac{4}{15} a\sigma^3 \delta q \nonumber \\
&& - a\left[ \frac{2}{3} \sigma^3 (\delta^3 - 3\delta^2 - \delta + 3) + m\sigma^2 (3\delta^2 - 6 \delta + 2)
+ 4 m^2 \sigma (\delta -1) + m^3\right]   .
\eea
The even gravitomagnetic and the odd gravitoelectric multipoles vanish identically because the solution possesses and additional reflection 
symmetry with 
respect to the hyperplane $y=0$ which can be interpreted as the equatorial plane.  
Higher odd gravitomagnetic and even gravitoelectric multipoles can be shown to be linearly dependent since they are completely determined in terms 
of the parameters $m$, $a$, $q$ and $\delta$. From the above expressions we see that the ZV parameter $\delta$  
enters explicitly the value
of the total mass $M_0$ as well as the angular momentum $J_1$ of the source. The mass quadrupole $M_2$ can be interpreted as a nonlinear superposition
of the quadrupoles corresponding to the ZV, ER and Kerr spacetimes. A generalization of the Kerr metric which includes an arbitrary quadrupole
moment is obtained by imposing the condition $\delta=1$. The resulting multipoles are
\begin{equation} 
M_{2k+1} = J_{2k}=0 \ ,  \quad k = 0,1,2,... 
\end{equation} 
\begin{equation} 
M_0 = m \ , \quad M_2 = - ma^2 + \frac{2}{15}qm^3 \left(1-\frac{a^2}{m^2}\right)^{3/2}  \ , ... 
\end{equation} 
\begin{equation} 
J_1= ma \ , \quad J_3 = -ma^3 +  \frac{4}{15}qm^3 a \left(1-\frac{a^2}{m^2}\right)^{3/2}  \ , ....
\end{equation}
It is interesting to note that this particular exact solution in the limit $a\rightarrow m$ leads to the spacetime of an extreme Kerr black hole,
regardless of the value of the quadrupole parameter $q$.

In all the above special solutions the electromagnetic field vanishes identically. One can easily obtain the corresponding electrovacuum 
generalizations by assuming that $e_0\neq 0$ and $g_0\neq 0$. The computation of the respective electromagnetic multipole moments can be
performed in an invariant manner and the result can be expressed as
\be
E_n = e_0 M_n \ , \quad H_n = g_0 J_n \ .
\ee
This means that the charge distribution resembles the mass distribution. The electric moments vanish identically if no mass distribution 
exists. This result is in accordance with our physical intuitive interpretation of a charge distribution. The magnetic moments turn out
to be proportional to the gravitomagnetic multipoles, with no magnetic monopole. This is a physical reasonable result in the sense that the 
magnetic field is generated by the motion of the charge distribution, i. e., in the present case, by the rotation of the compact object. 

It is worth noticing that in all the above special solutions 
we assumed that $\alpha_1=\alpha_2$ and obtained generalizations of the Kerr metric with arbitrary 
quadrupole moment. More general solutions can be obtained by relaxing this condition. Consider, for instance, the special solution
with $\delta=1$, $q_0=1$, $q_i=0$ for $i=1,2,...$, and
\be
\alpha_1 = \frac{\sigma-\frac{m}{\eta}}{a+\frac{l}{\eta}}\ ,\quad
\alpha_2 = \frac{\sigma-\frac{m}{\eta}}{a-\frac{l}{\eta}}\ ,\quad
%\ee
%with 
%\be 
\sigma^2 = \frac{m^2+l^2}{\eta^2} - a^2\ , \quad \eta = \frac{1}{\sqrt{1-e_0^2}} \ .
\ee
The resulting potential corresponds to  the charged Kerr-Taub-NUT spacetime with total charge $Q_0 = me_0$, where 
$l$ is the Taub-NUT parameter.  

The analysis of the general stationary solution turns out to be very complicated because of its mathematical complexity.
The special case of the Kerr metric with only an additional quadrupole parameter $q_2$ was recently analyzed in detail 
in \cite{bglq09}. It was shown that the presence of the quadrupole parameter completely changes the structure of spacetime
due especially to the fact that a naked singularity appears that affects the geometric structure of the ergosphere and
the motion of test particles around the mass distribution.  We can expect that similar effects will occur if higher 
multipole moments are taken into account.

%%%%%%%%%%%%%%%%%%%%%%%%%%%%%%%%%%%%%%%%%%%%%%%%%%%%%%%%%%%%%%%%%%%%%%%%%%%%%%%%%%%%%%
%%%%%%%%%%%%%%%%%%%%%%%%%%%%%%%%%%%%%%%%%%%%%%%%%%%%%%%%%%%%%%%%%%%%%%%%%%%%%%%%%%%%%%
\subsection{An interior solution}
\label{sec:interior}

\begin{figure}
%\begin{center}
\includegraphics[scale=0.4]{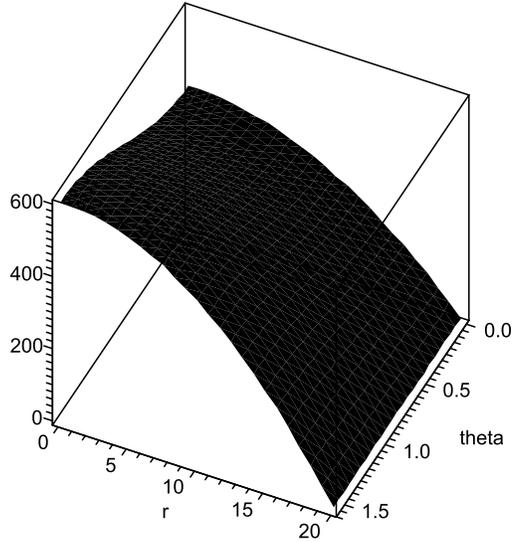}
%\end{center}
\caption{Plot of the inner pressure as a function of the spatial coordinates.
\label{fig4}
}
\end{figure}

A major problem in general relativity is to find physically meaningful solutions that can be matched with 
exterior exact solutions. In the context of multipolar solutions, one can say that only a few interior 
Schwarzschild solutions are known which can be considered as physically reasonable. The search for an interior 
solution that could be matched with the Kerr metric is still an open problem. Only recently it was proposed to
use the quadrupole moment as a parameter that introduces an additional degree of freedom into the differential 
equations which determine the internal structure of the mass distribution \cite{quev10b}. To illustrate the
method we consider the simplest generalization of the Schwarzschild spacetime with quadrupole moment given in 
Eq.(\ref{zv}). In the search for the corresponding interior solution, we found that an appropriate
form of the line element can be written as
\be
ds^2 = fdt^2 - \frac{e^{2k_0}}{f}\left(\frac{dr^2}{h} + d\theta^2\right) -\frac{\mu^2}{f}d\varphi^2\ ,
\ee
where 
\be
e^{2k_0} = (r^2-2mr+m^2\cos^2\theta)e^{2 k (r,\theta)}\ ,
\ee
and $f=f(r,\theta)$, $h=h(r)$, and $\mu=\mu(r,\theta)$. This line element preserves axial symmetry and 
staticity. 
In general, 
in order to solve Einstein's equations with a perfect fluid source, the  
pressure and the energy must be functions of the coordinates $r$ and $\theta$. 
However, if we assume that 
$\rho=$ const, the complexity of the corresponding differential equations reduces drastically:
\be
p_r = - \frac{1}{2} (p+\rho) \frac{f_r}{f}\ , \quad p_\theta =  - \frac{1}{2} (p+\rho) \frac{f_\theta}{f}\ ,
\ee
\be
\mu_{rr} = -\frac{1}{2h} \left( 2 \mu_{\theta\theta} + h_{r} \mu_r - 32 \pi p\frac{\mu e^{2\gamma_0}}{f} \right) \ ,
\ee
\be
f_{rr} = \frac{f_r^2}{f} -\left(\frac{h_r}{2h} + \frac{\mu_r}{\mu}\right)f_r + \frac{f_\theta^2}{hf} -
\frac{\mu_\theta f_\theta}{\mu h} -\frac{f_{\theta\theta}}{h} + 8\pi \frac{(3p+\rho)e^{2\gamma_0}}{h}\ .
\ee
Moreover, the function $k$ turns out to be determined by a set of two partial differential equations
which can be integrated by quadratures once $f$ and $\mu$ are known. The integrability condition of these partial
differential equations turns out to be satisfied identically by virtue of the remaining field equations.
It is then possible to perform a numerical integration by imposing
appropriate initial conditions. In particular, we demand that the metric functions and the pressure are finite 
at the axis. It turns out to be possible to find numerical solutions for the metric functions and the 
thermodynamic variables. In particular, the pressure behaves as shown 
in Fig.\ref{fig4}.

It can be seen that the pressure is finite in the entire interior domain, and tends to zero at certain hypersurface $R(r,\theta)$ 
which depends on the initial value of the pressure on the axis. Incidentally, it turns out that by increasing the value of the 
pressure on the axis, the ``radius function'' $R(r,\theta)$ can be reduced. Furthermore, if we demand that the hypersurface $R(r,\theta)$ 
coincides with the origin of coordinates, the value of the pressure at that point diverges. 
From a physical point of view, this is exactly the behavior that 
is expected from a physically meaningful  pressure function. 

This solution can be used to calculate numerically the corresponding Riemann tensor and its eigenvalues. As a result we obtain
that the solution is free of singularities in the entire region contained within the radius function $R(r,\theta)$. To perform the 
matching with the exterior metric (\ref{zv}) we applied the method formulated in \cite{quev10b} which uses the invariant 
properties of the eigenvalues of the curvature tensor.

\section{Conclusions}

We presented in this work a class of electrovacuum solutions of Einstein-Maxwell equations which can be used to describe 
the exterior gravitational field of a rotating distribution of mass endowed with an electromagnetic field. This class
of solutions is characterized by different sets of arbitrary parameters  which determine the multipolar structure of
the gravitational source. An important consequence of the presence of multipoles higher than the monopole is that 
the structure of the spacetime completely changes due to the appearance of naked singularities. In all the cases we 
investigated, the curvature singularities are situated inside the horizon which becomes the outermost naked singularity.  
This means that in principle it should be possible to find an interior solution that covers the entire spatial 
region where  the naked singularities exist. In particular, we found numerically an inner solution that can
be matched with an exact exterior metric with a particular quadrupole moment. The entire spacetime is shown to be 
free of singularities so that the entire manifold is well-defined in terms of solutions of Einstein's equations.
However, the problem of finding interior solutions taking into account the rotation of the gravitational source
remains an open problem. We believe that the multipolar structure of the solutions presented here could be used
as additional degrees of freedom to search for more realistic inner solutions. 

\section*{Acknowledgements}

%\begin{theacknowledgments}

I would like to thank the organizers of the XIV Brazilian School of Cosmology and Gravitation 
for the invitation to participate. This work was partially supported by DGAPA-UNAM, grant No. IN106110.

%\end{theacknowledgments}



\end{document}